\documentclass[oldversion]{aa} 
\usepackage{amsmath}
\usepackage{graphicx}
\usepackage{color}
\usepackage{txfonts}
\usepackage[authoryear]{natbib}

\begin{document}
\title{Continued decay in the cyclotron line energy in Hercules X-1} 
%\title{Long-term change in the cyclotron line energy in Hercules X-1} 

\author{R.~Staubert\inst{1}, D.~Klochkov\inst{1}, V.~Vybornov\inst{1},
  J.~Wilms\inst{2},  F.A.~Harrison\inst{3}}

\offprints{staubert@astro.uni-tuebingen.de}

\institute{
	Institut f\"ur Astronomie und Astrophysik, Universit\"at T\"ubingen,
	Sand 1, 72076 T\"ubingen, Germany
\and
        Dr.\ Remeis-Sternwarte, Astronomisches Institut der
	Universit\"at Erlangen-N\"urnberg, Sternwartstr. 7, 96049 Bamberg, Germany
\and
        Cahill Center for Astronomy and Astrophysics, California Institute of Technology, 
        Pasadena, CA 91125, USA 
}

\date{received: 2015 Dec 08, accepted: 2016 March 18}
\authorrunning{Staubert et al.}
\titlerunning{Continued decay in the cyclotron line energy in Her X-1}

\abstract
{The centroid energy $E_\mathrm{cyc}$ of the cyclotron line in the spectrum of the
  binary X-ray pulsar \object{Her~X-1} has been found to decrease with time on
  a time scale of a few tens of years - surprisingly short in astrophysical terms. 
 This was found for the pulse phase-averaged line centroid energy using 
  observational data from various X-ray satellites over the time period 
  1996 to 2012, establishing a reduction of $\sim$4\,keV. Here we
  report on the result of a new observation by \textsl{NuSTAR }
  performed in August 2015. The earlier results are confirmed and 
  strengthened with respect to both the dependence of $E_\mathrm{cyc}$
  on flux (it is still present after 2006) and the dependence on time:
  the long-term decay continued with the same rate,
  corresponding to a reduction of $\sim$5\,keV in 20 years.
}

   
   \maketitle
%
%_______________________________________________________________

\section{Introduction}
\label{sec:introduction}

This note is a continuation, an update, and an extension of 
\citet{Staubert_etal14}, hereafter ``Paper 1'',
which established the long-term change in the cyclotron line energy 
in Her~X-1. For the introduction to this source, a well-studied accreting 
X-ray binary pulsar, we therefore refer to Paper 1, 
except for brief remarks on the cyclotron line. The continuum spectrum
(a power law with exponential cutoff) is modified by a line-like feature.
This feature, discovered in 1976 in a balloon observation 
\citep{Truemper_etal78}, has been re-measured numerous times and is 
now generally accepted as an absorption feature around
40\,{\rm keV} due to resonant scattering of photons off electrons on
quantized energy levels (Landau levels) in the teragauss magnetic
field at the polar cap of the neutron star. The feature is therefore
also referred to as a \textsl{cyclotron resonant scattering feature} (CRSF).  
The energy spacing between the Landau levels is given by $E_\mathrm{cyc}$ =
$\hbar$eB/(m$_{\rm e}$c) = 11.6\,\text{keV}\,$B_\mathrm{12}$, where
$B_\mathrm{12}$=B/10$^{12}$\,\text{G}, providing a direct method of measuring
the magnetic field strength at the site of the emission of the X-ray spectrum. The 
observed line energy is subject to gravitational redshift $z$ at the location where 
the line is formed, such that the magnetic field can be estimated by
$B_\mathrm{12}$ = (1+z)~$E_\mathrm{obs}$/11.6\,${\rm keV}$, with $E_\mathrm{obs}$ 
being the observed cyclotron line energy.  The discovery of the
cyclotron feature in the spectrum of Her X-1 provided the first ever
direct measurement of the magnetic field strength of a neutron star,
in the sense that no other model assumptions are needed.  
Originally considered an exception, cyclotron features are now known to 
be rather common in accreting X-ray pulsars; $\sim25$ binary pulsars 
have now been confirmed as cyclotron line sources \citep{CaballeroWilms_12}. 

The centroid energy of the cyclotron line of Her~X-1 has been found to
vary systematically with respect to the following variables: \\
- Variation with phase of the 1.24\,s pulsation, with a peak-to-peak
amplitude of $\sim25$\% \citep{Voges_etal82,Vasco_etal13};\\
- Variation with X-ray luminosity, on both long and on short time
scales \citep{Staubert_etal07,Klochkov_etal11};\\ 
- Variation with phase of the 35\,d precessional period (albeit
rather weak) (see Paper 1); \\  %\citep{Staubert_etal14}.\\
- Variation with time: 1) a rather sharp jump upwards in energy around 1992
(\citealt{Gruber_etal01}, Paper 1), and
2) a true decrease in the phase-averaged line energy after 1993 with 
a magnitude of $\sim5$\,keV over 20 years 
(Paper 1; \citealt{Staubert_14,Klochkov_etal15}).

Here we present the results of a new measurement of the line energy
by NuSTAR, performed in August 2015. The earlier results presented in 
%\citet{Staubert_etal14}, 
Paper 1, \citet{Staubert_14}, and \citet{Klochkov_etal15} 
are confirmed and strengthened with respect to both the dependence 
of $E_\mathrm{cyc}$ on flux (it is still present after 2006) and the
dependence on time: the long-term decay continued at the same rate.
% of (-$0.264\pm0.014$)\,keV/yr.

% Fig. 1  new ----------------------------------------------------------------
\begin{figure}
\vspace{-1cm}
\includegraphics[angle=90,width=9.8cm]{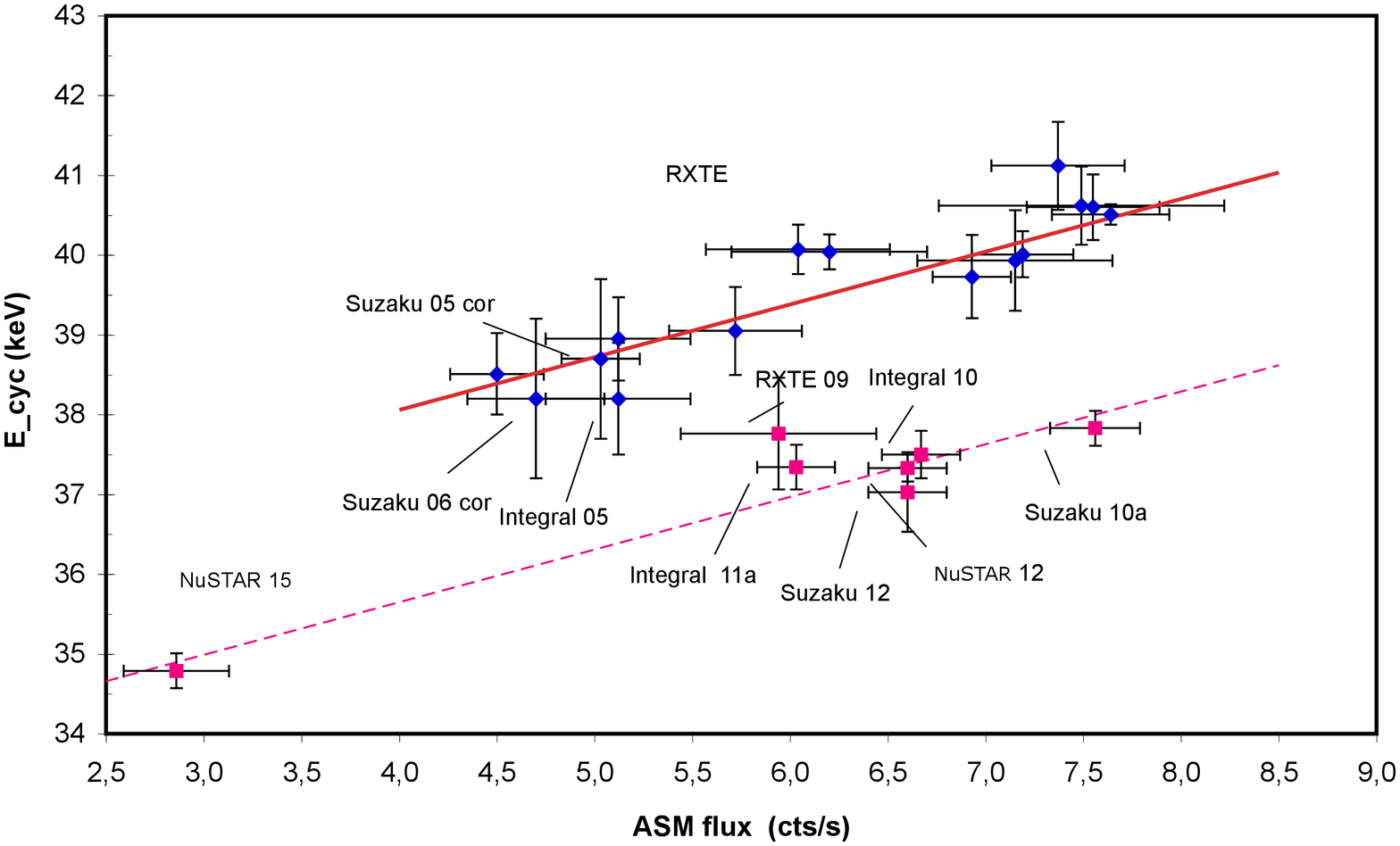}
\hfill
\vspace{-1.3cm}
\caption{Positive correlation between the cyclotron line energy 
  and the maximum X-ray flux of the corresponding 35-day cycle.
  Here, Fig.~1 of Paper 1 %\citet{Staubert_etal14} 
  is reproduced with the new \textsl{NuSTAR} point added: 
  $E_\mathrm{cyc}$(observed) = $34.79\pm0.22$\,keV. 
  The blue diamonds are values observed until 2006, the red dots are after 2006.
  The solid red line is a linear fit to data until 2006 with the original slope of
  0.66\,keV/(ASM cts/s), as found by \citet{Staubert_etal07}.
  The dotted red line is the best fit to the data after 2006 with the
  slope fixed to the same value. The flux level of the new
  observation (2.96 ASM-cts/s) is very low, extending the
  dynamic range of observed fluxes to the low end
  (missing for the time after 2006). The new value fits well
  into the general picture.
}
   \label{fig:correlation}
\end{figure}
%Fig. 1
%-------------------------------------------------------------------------------------

\vspace{-2mm}
\section{Observations}
\label{sec:data_base}

Her~X-1 was observed on August 3--4, 2015, in a campaign coordinated
by \textsl{NuSTAR} \citep{Harrison_etal13} and
\textsl{INTEGRAL} \citep{Winkler_etal03}.
For this contribution we restrict ourselves to the results of the first
13.5\,hr \textsl{NuSTAR} observation (ObsIDs 90102002002, 
MJD 57237.70 -- 57238.26) with a total on-source integration 
time of 27.1\,ksec. This is a Main-On observation of
35-day cycle 457 at phase 0.13\footnote{For cycle number and 
phase see \citet{Staubert_etal83,Staubert_etal13}}.
The mean \textsl{NuSTAR} count rate during this time was
$\sim43.5$\,cts/s (5-79\,keV). This count rate is very low for
Her~X-1, but turned out to be quite useful since it extended the dynamical
range of observed fluxes (in fact the lowest level at which the
cyclotron line energy has ever been measured). We note that for comparing 
\textsl{flux}, we use the maximum \textsl{Main-On} flux in units of 
\textsl{RXTE}/ASM-cts/s, as today it is determined through the
monitoring observations by \textsl{Swift}/BAT. The conversion
between these two units is as follows: 
(2-10\,keV ASM-cts/s) = 93.0 $\times$ (15-50\,keV BAT-cts~cm$^{-2}$~s$^{-1}$).
We determined the relationship by using flux measurements 
of times when both instruments operated simultaneously (see the Appendix).
The corresponding \textsl{Main-On-flux} for cycle 457 was 
$(2.96\pm0.28$)\,ASM-cts/s. 
We have verified that the maximum flux
of the \textsl{Main-On} is a good measure of the luminosity of the
source during the cycle \citep{Vasco_etal11}.

For this work we have neglected the point measured by \textsl{INTEGRAL} 
in 2012 because of the unexplained high value (see
discussion in Paper 1). %\citealt{Staubert_etal14}). 
A re-analysis of these data with the newest calibration, which 
will be contained in the new analysis software 
(OSA-11)\footnote{http://www.isdc.unige.ch/integral/analysis\#Software},
to be released later this year, has shown, that this data point
indeed needs to be corrected downward. A publication in which calibration 
issues and the evolution of the various software releases will be
discussed in depth is in preparation.
We also stress that the inclusion of this point (even with
the high value) in no way alters the fit results and 
the overall conclusions (the statistical weight is simply too low).
The results of the other observations from the
Agust 2015 campaign will be presented elsewhere.\footnote{The data from 
the simultaneous observations by \textsl{NuSTAR} and \textsl{INTEGRAL} 
will allow the respective inter-calibration to be studied again.}

\vspace{-2mm}
\section{Spectral results}
\label{sec:spectra}

%\vspace{-3mm}
 The spectral analysis was performed using the standard software as
part of HEASoft\footnote{http://heasarc.nasa.gov/lheasoft} (Version
6.16) provided by the \textsl{NuSTAR} team. 
\textsl{NuSTAR} employs two focussing X-ray telescopes
allowing observations in the 3--79\,keV energy range, each with its
own CdZnTe pixel detector \citep{Harrison_etal13}. The calibration
of these instruments is described in \citet{Madsen_etal15}.
For the Her~X-1 analysis, detector events from a region of 120 arcsec 
radius around the source position were used. A region at the periphery
of the field of view with a radius of 80 arcsec was used for background
determination (with 1.5$\cdot$10$^{-3}$  of the Her~X-1 flux, which is negligible).
Both telescopes/detectors were used to generate spectra in the 
5--79\,keV range. When combining these data for a common spectrum, 
a normalization factor was a free parameter.

For the spectral model we chose the \texttt{highecut}\footnote{
http://heasarc.nasa.gov/xanadu/xspec/manual/} %XSmodelHighecut.html}
model, which is based on a power law continuum with exponential cut-off;  
the CRSF is modeled by a multiplicative absorption line with a Gaussian 
optical depth profile. 
%Details of the fitting procedure can be found in the papers cited above. 
In order to smooth the jump in the derivative of the \texttt{highecut}
function at E$_{\rm cut}$, a multiplicative Gaussian was used, as in
\citet{Coburn_etal02}.
The spectral fit is good (reduced $\chi^{2}$=1.14 for 1777 degrees of freedom 
(dof)), yielding parameters that are very consistent with those found previously.             
The observed centroid energy of the CRSF is  
$E_\mathrm{cyc}$ = $34.79\pm0.22$\,keV. Throughout this work 
uncertainties quoted are at the 1$\sigma$ ($68\%$) level.

\vspace{-2mm}
\section{Variation of $E_\mathrm{cyc}$ with luminosity}
\label{sec:luminosity-dependence}

After the first observations of Her~X-1 by \textsl{RXTE} in 1996 and
1997 which yielded $E_\mathrm{cyc}$ values slightly lower than those of 
\textsl{CGRO}/BATSE and \textsl{Beppo}/SAX a few years earlier,
 we had suspected that there
might be a slow decay of $E_\mathrm{cyc}$ with time. This motivated
repeated and successful observing proposals over the following two decades.
In a series of \textsl{RXTE} observations until 2005, the apparent
decrease seemed to continue until this date. While working with a
uniform set of \textsl{RXTE} data between 1996 and 2005 trying to 
establish this decrease, we instead discovered that there was a 
dependence of $E_\mathrm{cyc}$ on X-ray flux \citep{Staubert_etal07}. 
This degraded the suspected decrease with time significantly, such
that it was neglected at this time.
The correlation found between $E_\mathrm{cyc}$ and
flux (luminosity) was positive, i.e., the cyclotron line 
energy $E_\mathrm{cyc}$ increases with increasing X-ray luminosity $L_\mathrm{X}$.
Her~X-1 was thus the first source that showed the opposite
behavior to those high luminosity transient sources in which
a negative correlation had already been seen in 1995 \citep{Mihara_95}.
Today we know more confirmed sources with a positive correlation
than with the negative one, see Sect.~\ref{sec:Discussion}.

Figure~\ref{fig:correlation} reproduces the correlation graph of 
Paper 1 (Fig.~1)
%\citet[][their Fig.1]{Staubert_etal14} 
with the new data point added (``NuSTAR 15''): 
the observed pulse phase averaged CRSF centroid energy is 
$E_\mathrm{cyc}$ = $34.79\pm0.22$\,keV. 
%(see Table~\ref{tab:new_results}). 
As noted in Paper 1,
%by \citet{Staubert_etal14}, 
the values after 2006 are
significantly lower than those before, which already clearly signals
a decrease in the cyclotron line energy with time.
In the earlier plot of 2014 the flux dependence for data after 2006
was less obvious. The new data point nicely confirms our previous
assumption that the same flux dependence prevailed after 2006
(dotted red line in Fig.~\ref{fig:correlation} is a fit through
the data after 2006 with the same slope as the solid red line).

%-----------------------------------------------------------------------------------------
%Fig. 2
\begin{figure*}
%\sidecaption
%\vspace{-1.5cm}
\includegraphics[width=14cm,angle=90]{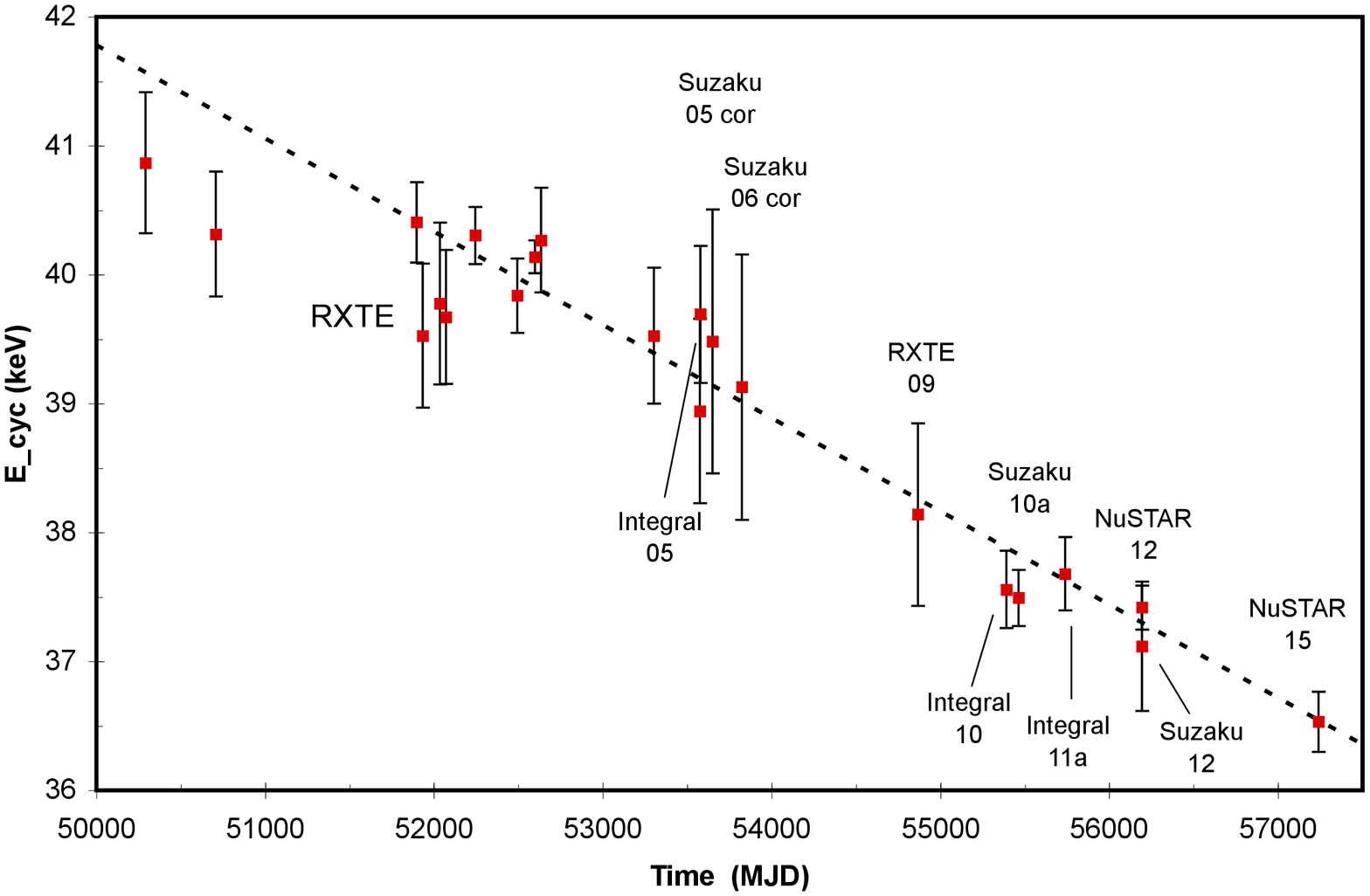}
\vspace{-2.5cm}
\caption{Her~X-1 pulse phase averaged cyclotron line energies
  $E_\mathrm{cyc}$ normalized to a reference ASM count rate of
  6.8\,cts/s using a flux dependence of 0.44\,keV/ASM-cts/s. 
  The new flux-normalized \textsl{NuSTAR} 2015 value is $36.49\pm0.24$\,keV.
  The best fit dashed line defines a linear decline of $E_\mathrm{cyc}$ with
  time with a slope of $-7.23\times 10^{-4}$\,keV$d^{-1}$.
 }
%\end{minipage}
 \label{fig:Fnorm}
\end{figure*}
%Fig. 2
%----------------------------------------------------------------------------------------

\vspace{-2mm}
\section{Variation of $E_\mathrm{cyc}$ with time: the long-term decay}
\label{sec:secular}

 The long-term decay of the pulse phase-averaged cyclotron line energy
$E_\mathrm{cyc}$ was finally established by \citet{Staubert_etal14} (Paper 1)
using data from different observatories taken between 1996 and 2012.
A linear reduction in $E_\mathrm{cyc}$ of $\sim$5\,keV over a time frame
of 20\,yrs was measured with high significance ($\sim18$ standard 
deviations). 

Here, we add the latest data point taken by \textsl{NuSTAR} in August
2015. 
First, we consider the time evolution of the flux-normalized
$E_\mathrm{cyc}$ for two different time periods: 1996 -- 2006 and
2006 -- 2015 (separated at MJD $\sim$54000) . The normalization 
to the reference flux of 6.8 ASM-cts/s was done using a slope of 
0.44\,keV per ASM-cts/s, which is the best fit slope from the 
simultaneous flux- and time-dependence of the previous data set. 
The slopes for those two time periods are 
$(-3.2\pm0.7)\times10^{-4}$\,keV/day and 
$(-7.0\pm0.4)\times 10^{-4}$\,keV/day, respectively,
constituting a significant steepening in the later period.

 We then repeat the simultaneous fit with two variables 
(the dependence on X-ray flux and the dependence on time) as in Paper 1,
with the \textsl{NuSTAR} 2015 point added, such that the complete data 
set of 1996 -- 2015 is used.\footnote{All points in this 
data set are from observations taken during the \textsl{Main-On} of 
Her~X-1 at 35\,d phases less than 0.2.}
 In order to separate the two variables, the following function was
 used: 
%$\begin{equation}  
%$E_\mathrm{cyc}$ (calc) = $E_\mathrm{0}$ + $a~\times$ ($F - F_\mathrm{0}$) + $b~\times$ ($T - T_\mathrm{0}$)\\
%\end{equation}
\begin{equation}  
E_\mathrm{cyc}(\mathrm{calc})=E_0 + a\cdot(F-F_0) + b\cdot(T-T_0)
%\{\rm E}_{\rm cyc}(\rm calc) = E_{\rm 0} + a \times (F - F_{0}) + b \times (T - T_{0}) \\
\label{equ:lin-lin}
\end{equation}
with $F$ being the X-ray flux (the maximum flux of the respective 35d cycle) in units of 
ASM-cts/s, as observed by \textsl{RXTE}/ASM (and/or \textsl{Swift}/BAT), with  
$F_\mathrm{0}$ = 6.80 ASM-cts/s, and $T$ being time in MJD with $T_\mathrm{0}$ = 53500.
The bi-linear fit is a good description of the complete 1996--2015 data set. 
The fit parameters are given in Table~\ref{tab:3D_3}.
A comparison  of the fit parameters with those of fit No.~4 in Table~4 of Paper 1,
%\citet{Staubert_etal14}, 
shows that all parameters are unchanged (only the uncertainty of parameter 
``a'' describing the flux dependence is slightly smaller now). 

%Table 1 (FTnorm3)-------------------------------------------------------------------------
\begin{table}
\caption[]{Parameters of the bi-linear fit with Equation~(1)  to $E_\mathrm{cyc}$ 
 values observed between 1996 and 2015 (excluding \textsl{INTEGRAL}12). 
%  without and with the extension by a quadratic term in the time dependence. 
The reference flux is F$_{0}$ = 6.8\,(ASM-cts/s) and the reference
time is T$_{0}$ = MJD 53500.}
%    \label{tab:3D_3}
\vspace{-3mm}
\begin{center}
\begin{tabular}{lllll}
\hline\noalign{\smallskip}
%\vspace{-1mm}
E$_{\rm 0}$             & $a$                          & $b$    & $\chi^{2}$ & dof      \\
 $[keV]$                & [keV/ASM-cts/s] & [$10^{-4}$ keV/d]   &              &    \\
\hline\noalign{\smallskip}
 $39.25\pm0.07$  & $0.44\pm0.06$  &  $-7.23\pm0.39$  &  20.4   &  19  \\   
   \noalign{\smallskip}\hline
   \label{tab:3D_3}
\end{tabular}\end{center}

\end{table}
%Table 1 -----------------------------------------------------------------------------------

Equation~(1) constitutes a plane in 3D
space defined by the three quantities ASM-flux, time, and $E_\mathrm{cyc}$.
In principle, the situation could be displayed by a 3D plot. Such a plot 
(also including the data points), however, is not easy to read in practice. 
We therefore show two corresponding 2D plots that make
use of normalized values.

1) We normalized the observed $E_\mathrm{cyc}$ values to the 
reference flux of 6.8\,(ASM-cts/s) using the flux 
dependence found in the simultaneous fit and plotted it against time.
Figure~\ref{fig:Fnorm} shows the remaining linear time dependence of
$E_\mathrm{cyc}$, demonstrating the long-term decay of $E_\mathrm{cyc}$
with a slope of  $(-7.23\pm0.39)\times 10^{-4}$\,keV/day
(or $-0.26\pm0.014$\,keV/yr) . The dashed line
is the linear best fit. The new data point (``NuSTAR 15'') is very
consistent with the earlier data.

2) We normalize the observed $E_\mathrm{cyc}$ values to the 
reference time T$_{0}$ = MJD 53500 using the time   
dependence found in the simultaneous fit and plotte against flux.
Figure~\ref{fig:Tnorm} shows the remaining linear flux dependence of 
$E_\mathrm{cyc}$.
The slope is $0.44\pm0.06$\,keV/(ASM-cts/s), which is the 
same as in Paper~1, but slightly lower
than the value first found in the discovery paper of \citet{Staubert_etal07}, 
where only data until 2005 and no time dependence were considered.

After including the 2015 data point of \textsl{NuSTAR}, the fact noted 
in Paper 1
%by \cite{Staubert_etal14} 
that the fit is slightly better when an additional quadratic term 
%c \times (T - T_{0})^{2}     $\chi^{2}
($c~(T-T_0)^{2}$) in the time dependence is introduced, continues to be valid:
%$E_\mathrm{0}$ = $39.41\pm0.11$\,keV, a
E$_{0}$ = $(39.41\pm0.11)$\,keV,  $a$ = $(0.38\pm0.07)$\,keV/ASM-cts/s,
$b$ = $-(6.68\pm0.48)~10^{-4}$\,keV/d, and 
$c$ = $-(5.3\pm2.8)~10^{-8}$\,keV/d$^{2}$, with $\chi^{2}$ = 16.7 for 
18 degrees of freedom.
The improvement in $\chi^{2}$ with respect to the purely linear fit
is, however, marginal; an F-test gives a probability of  6.1\%  
for the improvement in $\chi^{2}$ being by chance when introducing the 
quadratic term.

If we determine the decay rate of $E_\mathrm{cyc}$ in the very recent time
frame, using only the two values measured by \textsl{NuSTAR} in 2012 
and 2015, we find $(-8.5\pm0.3)\times 10^{-4}$\,keV/day. We note that,
owing to the very accurate measurements by \textsl{NuSTAR}, the
uncertainty is quite small (similar to that of the fit to the complete
data).

We also refer to the independent analysis of spectral measurements
by \textsl{Swift}/BAT, which confirms the decay of $E_\mathrm{cyc}$ over
the time frame 2005--2014 \citep{Klochkov_etal15}. In this work each 
data point is an average over several 35-day cycles, neglecting 
(in a sense averaging over) the flux dependence of the cyclotron line 
energy. Adding these points (which generally have larger uncertainties)
to the previous data set, we find that they closely match the
earlier picture and the parameters of the linear fit (simultaneous in
flux and time) are not changed. However, owing to the wider spread of
the BAT data, the reduced $\chi^{2}$ is significantly increased (from 
$\sim 1$ to beyond 3).

With reference to the Appendix and Fig.~\ref{fig:figA2} we
need to modify the statement made in Paper~1 that the mean maximum 
X-ray flux (when averaged over several 35d cycles) is constant.
A new analysis has now found that the data of the last 20\,yrs 
formally define a slight flux reduction of $0.95\pm0.07$ ASM-cts/s.
We note, however, that the scattering of the individual maximum flux
measurements is on the same order of magnitude, the observed 
variation over several 35d cycles can even exceed 100\% 
(a factor of two).

\vspace{-2mm}
\section{Summary of observational results}
\label{sec:Summary}

The current situation regarding the variation of the phase-averaged 
cyclotron line energy with luminosity and with time
is the following:
\begin{enumerate}
\item The latest measurement of  $E_\mathrm{cyc}$ confirms that 
the long-term decay of the cyclotron line energy continues as before, 
with a rate of $-(0.260\pm0.014)$\,keV/yr (see Fig.~\ref{fig:Fnorm}).
An independent confirmation of the decay is provided by
\textsl{Swift}/BAT observations for the time interval 2005--2014
\citep{Klochkov_etal15}. 
This work provides an idea about the evolution
of the originally measured (not flux normalized) $E_\mathrm{cyc}$
values, in a similar way to that in Fig.~4 of Paper~1.

\item The dependency of $E_\mathrm{cyc}$ on flux (or luminosity)
as detected by \cite{Staubert_etal07} is still valid after 2006 (see 
Fig.~\ref{fig:correlation} and Fig.~\ref{fig:Tnorm}). In the combined flux- and 
time-dependent analysis of the data available until 2012 (Paper 1)
%\citep{Staubert_etal14} 
this was an assumption that has now been nicely confirmed (with 
the help of the accidentally low flux during the last observation - even 
though regrettable from the statistics point of view).

We would like to add another remark related to the unusually low
flux level that we encountered during the 2015 \textsl{NuSTAR} 
observation. We have since realized that Her~X-1 was on its way 
to a new \textsl{anomalous low} that was reached around 
MJD 57340. Such lows happen irregularly on a time scale of 
$\sim$5\,yrs or $\sim$10\,yrs. It is a known feature that 
the flux fades gradually when going into such a low 
\citep{Coburn_etal00}. We would like to point out that the perfect
correlation between flux and $E_\mathrm{cyc}$ 
(Fig.~\ref{fig:Tnorm}) indicates a true reduction of luminosity 
(accretion rate) in the source, and not a reduction in observable
flux by a progressing shading of the X-ray emitting region by 
the accretion disk. This view is further supported by the fact
that the entrance into a new \textsl{anomalous low} coincides
with a strong spin-down\footnote{The Gamma Ray Burst Monitor 
(GBM) on \textsl{Fermi} provides pulse frequencies: 
http:/gammaray.msfc.nasa.gov/gbm/science/pulsars.html}
(owing to a reduced mass accretion rate).

%-----------------------------------------------------------------------------------------
%Fig. 3
\begin{figure}
\vspace{-0.5cm}
\includegraphics[angle=90,width=9.8cm]{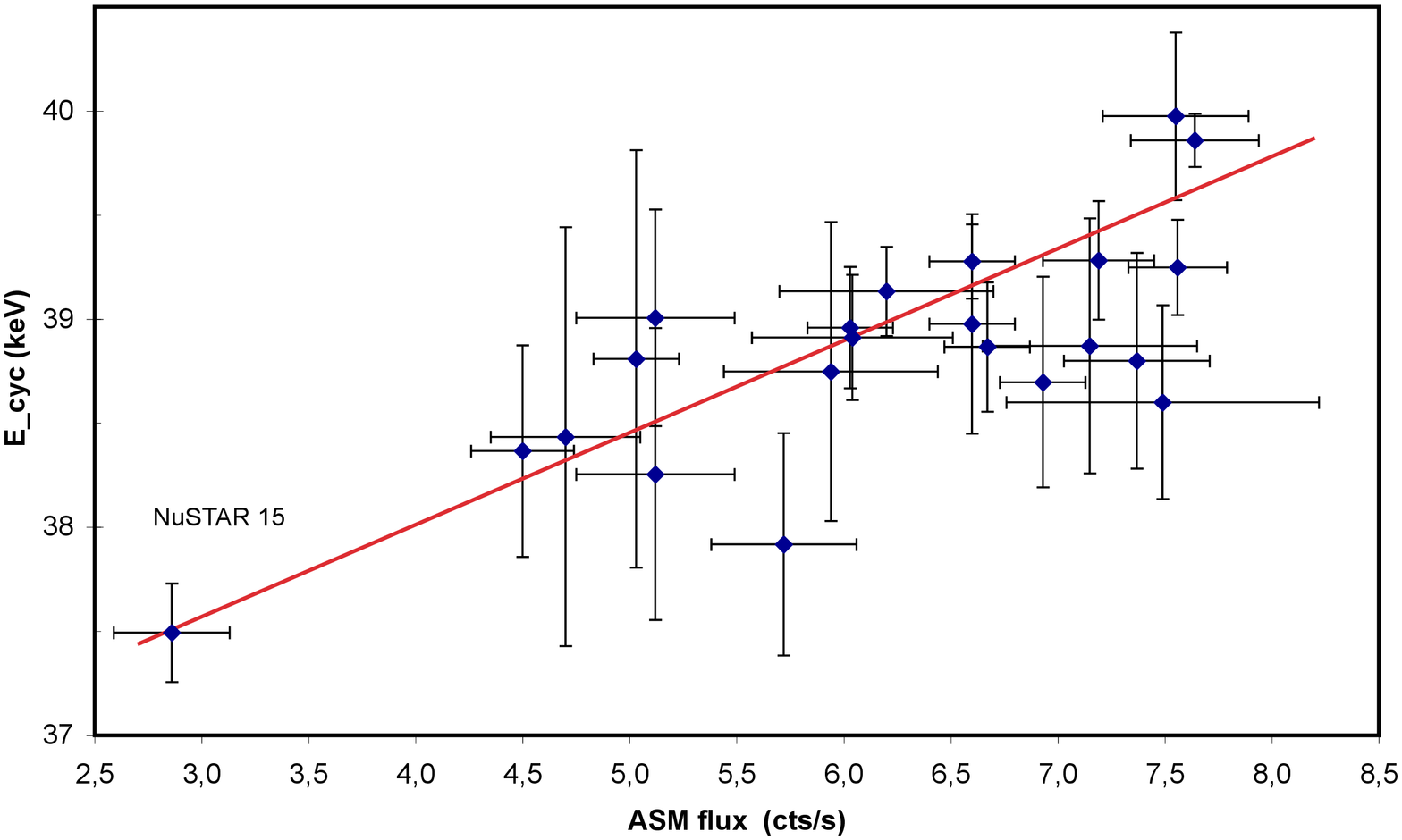}
\hfill
%\begin{minipage}[b]{1.0\textwidth}
\vspace{-1cm}
\caption{Her~X-1 pulse-phase-averaged cyclotron line energies
  $E_\mathrm{cyc}$ normalized to the reference time MJD 53500 using a
  time dependence of  $-7.22\times 10^{-4}$\,keV$d^{-1}$.
  The new time-normalized \textsl{NuSTAR} 2015 value is $37.49\pm0.24$\,keV}.
 The solid line defines the linear best fit for the flux dependence of $E_\mathrm{cyc}$
  with a slope of 0.44\,keV/(ASM-cts/s).
%\end{minipage}
 \label{fig:Tnorm}
\end{figure}
%Fig. 3
%----------------------------------------------------------------------------------------

\item It is conceivable that the decay of $E_\mathrm{cyc}$ is actually
accelerating with time. Even though the combined fit with a quadratic 
term in the time dependence yields only marginal evidence for a true (negative) 
quadratic term, a change in the decay rate is supported by the following two facts. 
First, using flux-normalized values, we find that the decay of 
$E_\mathrm{cyc}$ is stronger during the more recent period 2006--2015
than during 1996--2006 (see Section~5). 
This steepening is highly significant, and is in line with the remark
already made in Paper 1
%by \citet{Staubert_etal14}
that in the earlier period the flux dependence is the dominant
effect, while in the later period it is the time dependence.
Second, using the two observations by \textsl{NuSTAR} in 2012 and
2015 alone, a decay rate of $(-8.5\pm0.3)\times 10^{-4}$\,keV/day is found, 
which is slightly steeper than the slope for the overall decay from 
1996--2015 (see Table~\ref{tab:3D_3}).

\item The slight flux reduction of $\sim$15\% 
between 1996 and 2016 (see Fig.~\ref{fig:figA2} in the Appendix) 
corresponds to a reduction in $E_\mathrm{cyc}$ by $\sim$0.4\,keV, 
when the found flux dependence of -0.44\,keV/(ASM-cts/s)
(see Table~\ref{tab:3D_3}) is applied. We emphasize that this
in no way interferes with the reduction of $E_\mathrm{cyc}$ 
with time (by $\sim$5\,keV over 20 yrs), since the long-term 
flux variation is also taken into account when the two-variable fit
(Eq.~\ref{equ:lin-lin}) is applied.

\end{enumerate}

\section{Discussion}
\label{sec:Discussion}

 How common are the two dependencies of $E_\mathrm{cyc}$ - on luminosity 
and on time - among binary X-ray pulsars? With regard to the dependence on
luminosity, the negative correlation (a decrease in $E_\mathrm{cyc}$ with
increasing  $L_\mathrm{X}$  ), was actually discovered first: in high luminosity
transient sources from observations by \textsl{Ginga} \citep{Mihara_95}.
However, of the three sources originally quoted (V~0332+53, 4U~0115+63,
and Cep~X-4) only one, V~0332+53, can today be considered a secure
source.\footnote{http://users.ph.tum.de/ga24wax/2015\_W6\_slides/Staubert\_Talk.pdf}
In Cep~X-4 the effect was never confirmed, and in 4U~0115+63, although
apparently re-measured 
\citep{Tsygankov_etal06,Nakajima_etal06,Tsygankov_etal07,Klochkov_etal11},
\citet{SMueller_etal13} have shown that the anti-correlations are most
likely an artifact introduced by the way the continuum was modeled
(see also \citet{Iyer_etal15}).

To the contrary, the positive correlation was only discovered in 2007 by 
\citet{Staubert_etal07} in Her~X-1, a persistent medium luminosity 
source. Since then 
four more sources (all at moderate to low luminosities) 
have been found with $E_\mathrm{cyc}$ increasing with increasing 
luminosity: GX~304-1 \citep{Yamamoto_etal11,Klochkov_etal12},
Vela~X-1 \citep{Fuerst_etal14}, A~0535+26 
\citep{Klochkov_etal11,Sartore_etal15}  %DMueller_etal13,
and Cep~X-4 \citep{Fuerst_etal15}. 
As noted above, Cep~X-4 is among those sources that were originally
considered to show a negative corelation \citep{Mihara_95}, even 
though the evidence for this source was not very strong and it was 
never confirmed. It would, however, be quite interesting to find  
both behaviors (positive and negative correlation) in the same source 
at very different luminosities. The postive $E_\mathrm{cyc}$/$L_\mathrm{X}$
correlation might well be a general property of binary X-ray pulsars
at low to medium luminosties.

With regard to the long-term change of the cyclotron line energy,
Her~X-1 is the first and still the only object, for which this 
phenomenon has been securely established.\footnote{
There are indications, though still fairly weak, for the opposite
behavior in 4U~1538$-$522 \citep{Hemphill_etal14}}.

For the physical interpretation of the observed phenomena,
$E_\mathrm{cyc}$/$L_\mathrm{X}$   correlation and the long-term change of $E_\mathrm{cyc}$,
we refer to the discussions in Paper 1
%\citet{Staubert_etal14} 
and \cite{Staubert_14}. Here we re-iterate only the following general ideas.

We assume that we can distinguish between two accretion regimes:
 $L_\mathrm{X}$$>$$L_\mathrm{crit}$ (super-critical) and
 $L_\mathrm{X}$$<$$L_\mathrm{crit}$ (sub-critical).
The early explanation for the decrease of $E_\mathrm{cyc}$ with increasing
 $L_\mathrm{X}$  at super-critical accretion is still popular: based on 
\citet{BaskoSunyaev_76}, who had shown that the height of the 
radiative shock above the neutron star surface should grow linearly
with increasing accretion rate, \citet{Burnard_etal91} noted
that this means a reduction in field strength and therefore in 
$E_\mathrm{cyc}$. For the positive $E_\mathrm{cyc}$/$L_\mathrm{X}$   correlation
(first seen in Her~X-1), \citet{Staubert_etal07} had proposed that
at sub-critical accretion the stopping mechanism of the accreted
material is not photon pressure, but is due instead to Coulomb
interactions \citep{Nelson_etal93}, which should lead to the
opposite behavior (see also \citealt{Becker_etal12}).
An alternative explanation, not connected to the height of
line emitting region, is to assume changes in the B-field 
configuration (orginally assumed to be a dipole field),
suffering distortions due to loading with a variable amount
of accreted material \citep{MukhBhatt_12}.
Recently, a new idea has been put forward by 
\citet{Mushtukov_etal15}, suggesting that a variation in accretion
rate should lead to a variation in the velocity distribution of
the infalling material in which the cyclotron line is assumed
to be generated, such that a variable Doppler effect is 
responsible for a shift in the observed line energy.

Regarding the long-term change of the cyclotron line energy,
we repeat here our earlier suggestion \citep{Staubert_14}
that a slight imbalance between the rate of accretion and the
rate of ``losing'' material at the bottom of the accretion
mound, either by incorporation into the neutron star crust
or leaking of material to larger areas of the neutron star surface
\citep{Mukherjee_etal13b,Mukherjee_etal14},
can lead to a change in the mass loading and consequently in the
structure of the accretion mound (height or B-field configuration).

%\vspace{-1mm}
\begin{acknowledgements}
The motivation for this paper are new observational data taken by the 
NASA satellite \textsl{NuSTAR}.  We would like to acknowledge the dedication 
of all the people who have contributed to the great success of this
mission, here especially Karl Forster for his effort with respect to the 
non-standard scheduling of the observations. Further important data
were provided by the equally successful missions \textsl{RXTE} 
and \textsl{Swift} of NASA and \textsl{INTEGRAL} of ESA.
This work was supported by the Deutsche Forschungsgemeinschaft (DFG) 
through joint grants KL~2734/2-1 and WI~1860/11-1.
We thank the anonymous referee for useful comments and suggestions.
\end{acknowledgements}

\vspace{-4mm}
\bibliographystyle{aa}
%\bibliography{accretion,her_oct06,a0535_sep06,v0332_sep06,gx301_sep06}
\bibliography{refs_herx1}

\newpage

\appendix
\section{Determination of X-ray flux and ASM~/~BAT intercalibration}

Here we provide technical details on the procedures used to determine the X-ray flux
as well as the intercalibration between \textsl{RXTE}/ASM and \textsl{Swift}/BAT.
Since the start of the operation of the \textsl{All Sky Monitor}
(ASM) \citep{Levine_etal96} on board the NASA satellite \textsl{RXTE} in 1996, 
a high quality continuous monitoring of sufficiently strong X-ray
sources has been underway. ASM provides average count rates in the 2--10\,keV
range (in units of cts/s) for 90 sec long ``dwells'' and for one 
day\footnote{http://xte.mit.edu}. The operation of ASM ended in early 2010.
Since early 2005 the \textsl{Burst Alert Telescope} (BAT)
\citep{Barthelmy_etal05} on board the NASA satellite \textsl{Swift} monitors 
X-ray sources (in addition to detecting gamma-ray bursts). The BAT average
count rates are given for the 15--50\,keV range in units of
cts/(cm$^{2}$~s), both per satellite orbit and 
per day\footnote{http://swift.gsfc.nasa.gov/results/transients}.
Between 2005 and 2010 both monitors operated simultanously, allowing
an inter-calibration of the two instruments to be established.

\subsection{Determination of X-ray flux}

\begin{figure}
%\sidecaption
\vspace{-1.0cm}
\includegraphics[width=7.2cm,angle=90]{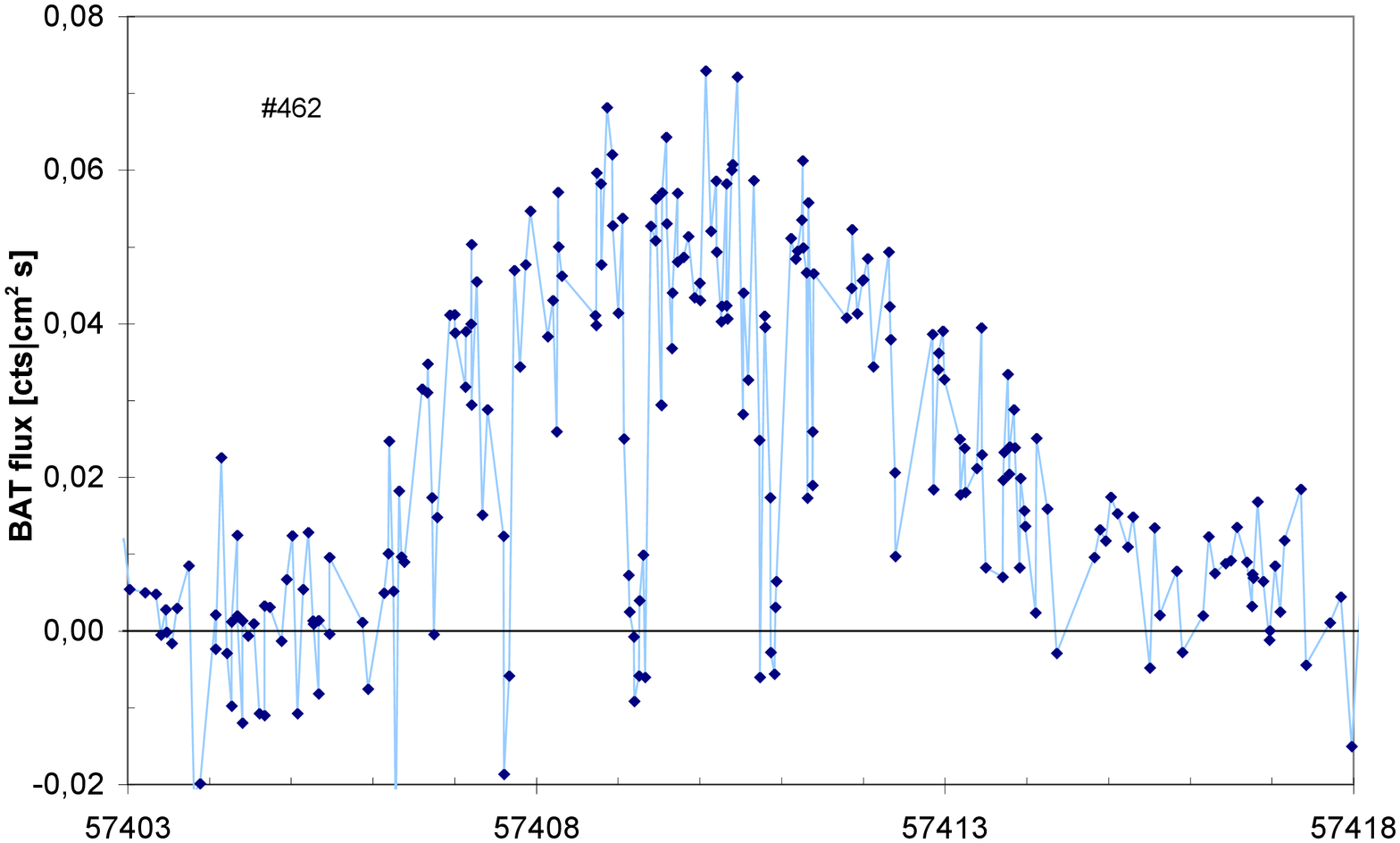}
\includegraphics[width=7.2cm,angle=90]{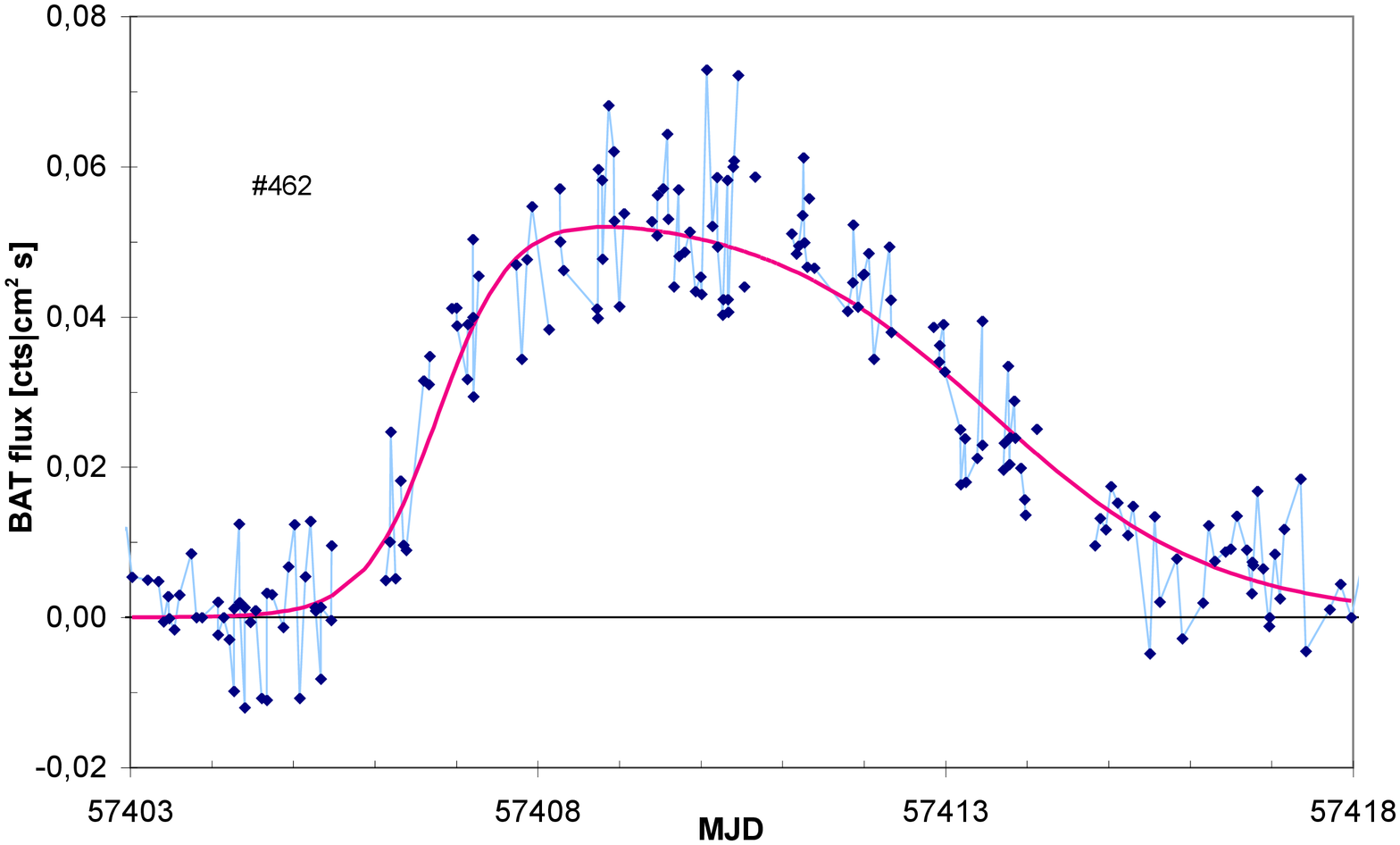}
%\includegraphics[width=10.5cm,angle=90]{plot_incl_NuS15_MJD.ps}
%\includegraphics[width=0.65\textwidth,angle=90]{plot_incl_NuS15_MJD.ps}
%\hfill
%\begin{minipage}[b]{0.3\textwidth}
\vspace{-1.0cm}
\caption{Her~X-1 light curve for the \textsl{Main-On} of 35d cycle 462, representing
  orbital averages as observed by \textsl{Swift}/BAT. Upper panel: observed 
  flux values. Lower panel: \textsl{cleaned} data with points during binary 
  eclipses and dips, and extreme outliers, removed.
  Uncertainties of individual measurements around the maximum flux are
  of the order of 20\%. The solid red line is the best fit function as
  defined in the text (Eq.~\ref{equ:main-on-fit}). The maximum flux
  for this 35d cycle is $0.052\pm0.002$ BAT-cts/(cm$^{2}$~s).
}
%\end{minipage}
 \label{fig:A1}
\end{figure}

As stated in the main text, we take the maximum flux 
encountered in each \textsl{Main-On} as a measure of the local
luminosity of this 35d cycle. How is this flux determined?
Since the discovery of Her~X-1 as an accreting binary X-ray pulsar by
\textsl{UHURU} in 1972 \citep{Tananbaum_etal72}, it has been known that
the flux varies strongly on three different time scales (in addition
to the 1.237\,s pulsation): the $\sim$35\,d on-off period, the 1.70\,d 
binary eclipses (of $\sim$6\,hours duration) and the 1.62\,d period with 
similarly long ``dips''.  For details on these modulations and the
underlying physics, see e.g., \citet{Klochkov_etal06} and \citet{Staubert_etal13}.
Fig.~\ref{fig:A1} (upper panel) shows the Her~X-1 light curve of the 
\textsl{Main-On} of 35d cycle~462, as measured (in orbital averages)
by \textsl{Swift}/BAT. All three types of flux modulations are
apparent. In order to find a measure of the true intrinsic luminosity
(accretion rate), the observed light curves were \textsl{cleaned} by
removing data points from times of the 
eclipses\footnote{For the binary ephemeris of Her~X-1 see \citealt{Staubert_etal09}.} 
and the dips, and also from some other
extreme outliers. The cleaned light curve of cycle~462 is
shown in the lower panel of Fig.~\ref{fig:A1}. Such cleaning was
performed and two characteristic parameters were extracted interactively 
for all 35d \textsl{Main-Ons}  observed by ASM and by BAT, the maximum
flux and the \textsl{turn-on} time. For the maximum flux an interval 
(of $\sim$3\,d) around the maximum was selected and the formal average 
(and its corresponding uncertainty) of all data points within this interval
determined. The turn-on time was found by taking the center between
the time of the first significant flux measurement (at the rise from
the off-state) and the time of the previous data point. The
corresponding uncertainty is half of the difference between those 
two times. 

%-----------------------------------------------------------------------------------------
%Fig. A1
%Fig. A1
%----------------------------------------------------------------------------------------

In addition to the above procedure, a second method was
applied for a limited number of 35d cycles, namely a formal best fit
to the cleaned light curve using the following function:
\begin{equation}  
%{\rm F}(\rm t) = A_0 * [1/(1+exp((A_1-t)/A_2))] *  [1/(1+exp((t-A_3)/A_4))]
{\rm F}(\rm t) = A_0 \times \frac{1}{1+exp(\frac{A_1-t}{A_2})} \times \frac{1}{1+exp(\frac{t-A_3}{A_4})}   
\label{equ:main-on-fit}
\end{equation}
This second method yielded results that were in good agreement with
the former procedure. The solid red line in Fig.~\ref{fig:A1} is the best 
fit to the cleaned light curve of 35d cycle 462.

%\begin{equation}  
%  [1/(1+exp((A_1-t)/A_2)))] *  [1/(1+exp((t-A_3)/A_4)))]
%\label{equ:main-on-fit}
%\end{equation}
%-----------------------------------------------------------------------------------------
%Fig. A2
\begin{figure*}
%\sidecaption
%\vspace{-1.5cm}
\includegraphics[width=14cm,angle=90]{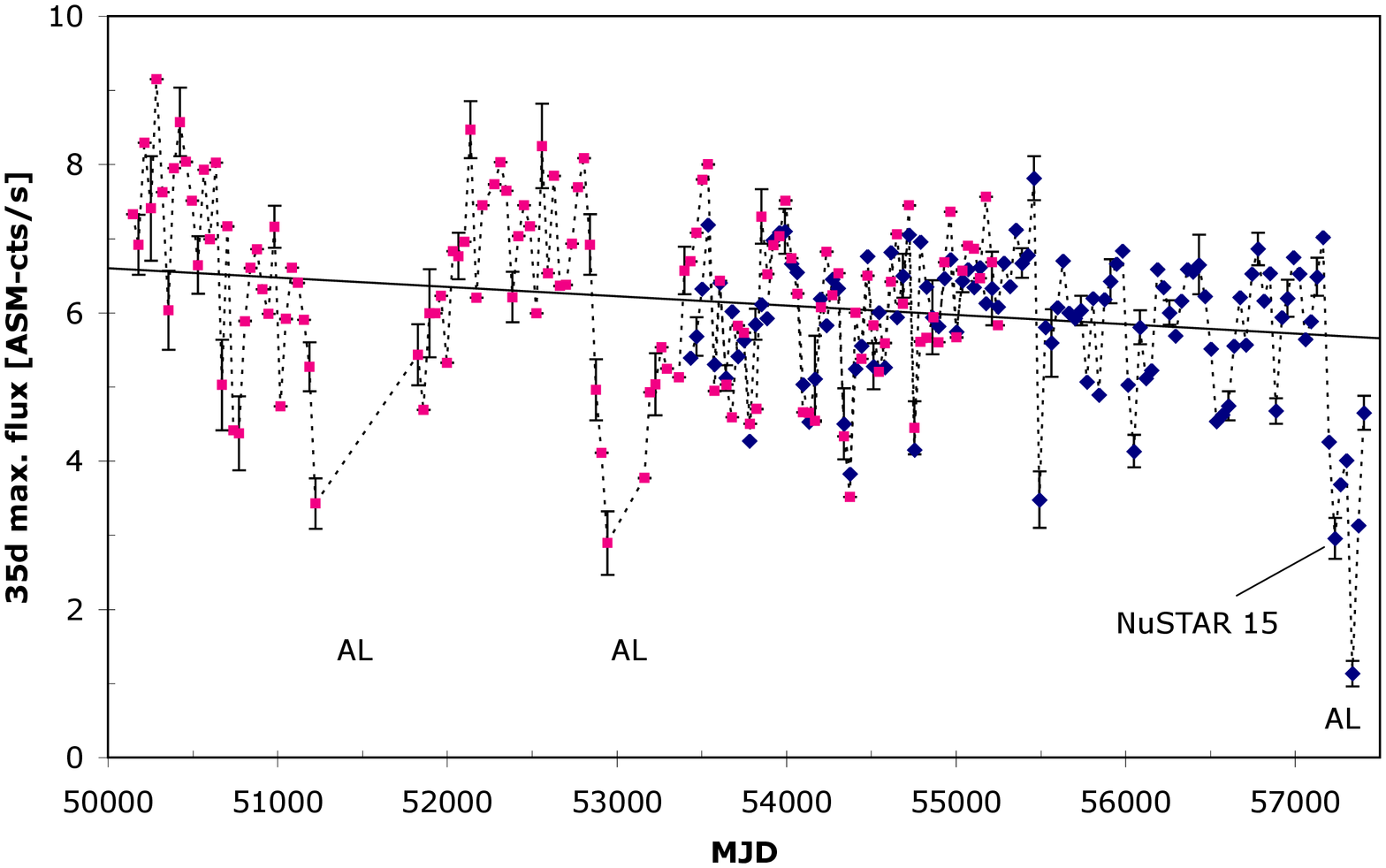}
%\includegraphics[width=14cm,angle=90]{ASM_BAT_max_lc.ps}
%\hfill
%\begin{minipage}[b]{0.3\textwidth}
\vspace{-2cm}
\caption{Her~X-1 long-term light curve, showing the maximum flux of
  each 35d cycle in units of ASM-cts/s, as determined from observations
  by the monitoring instruments: red points are from \textsl{RXTE}/ASM,
  blue points are from \textsl{Swift}/BAT. Measured BAT values (in units
  of cts/(cm$^{2}$~s)) were scaled to ASM-cts/s using the scaling factor
  determined from the overlapping time period (MJD 53433--55248).
  For better readability of the light curve, only a few error bars are
  shown. Three \textsl{anomalous lows} \citep{Coburn_etal00} can be 
  recognized (indicated by AL). The rather low flux value of the 
  latest \textsl{NuSTAR} observation is also marked (NuSTAR 15).
  The solid line is a linear fit to the data (excluding those points
  belonging to the ALs). Its slope is -$(1.30\pm0.09)$~10$^{-4}$ 
  (ASM-cts/s)/d, which corresponds to a small reduction in flux of
  ($0.95\pm0.07$) ASM-cts/s over 20 years.
}
%\end{minipage}
 \label{fig:figA2}
\end{figure*}
%Fig. A2
%----------------------------------------------------------------------------------------

\subsection{Intercalibration between ASM and BAT}

To be able to use flux measurements of the two monitoring instruments
\textsl{RXTE}/ASM and \textsl{Swift}/BAT in a common analysis, an
intercalibration was performed. We used the overlapping time
period of about 5\,yrs (2005--2010) for which both instruments
operated simultaneously. Fig.~\ref{fig:figA2} shows a common light
curve of Her~X-1 from 1997 to 2016, showing the maximum 
\textsl{Main-On} flux (see previous section). For this light curve the
observed BAT maximum values were multiplied with the scaling 
factor \\
(2-10\,keV ASM-cts/s) = 93.0 $\times$ (15-50\,keV BAT-cts~cm$^{-2}$~s$^{-1}$). \\

This scaling factor was found by plotting the measured maximum
values of both monitors against each other: a linear fit to those values
yielded the conversion factor of $93.0\pm1.3$. As a second method,
we used the measured one-day averages from both monitors: the
correlation yielded the same result. In order to test, whether the
conversion factor showed any time dependence, we divided the 
overlapping period into five intervals and determined the individual
conversion factors. The results are summarized in
Table~\ref{tab:tableA2}. We conclude that the values are consistent
with a constant conversion factor.
When producing frequency distributions of the maximum flux values of
Fig.~\ref{fig:figA2} (over a time period of a few years), we find that
they are all consistent with
Gaussian distributions with a standard deviation of $\sim$1 ASM-cts/s
(around 15\% of the mean flux). The variation over several 35d
cycles, however, can exceed 100\%.

%Table A2-------------------------------------------------------------------------
\begin{table}
\caption[]{Conversion factor f between fluxes measured with ASM and
  BAT for five shorter time intervals and for the total overlapping
  time:\\ 
(2-10\,keV ASM-cts~s$^{-1}$) = f $\times$ (15-50\,keV BAT-cts~cm$^{-2}$~s$^{-1}$).
}
%\vspace{-3mm}
\begin{center}
\begin{tabular}{lllll}
\hline\noalign{\smallskip}
Time interval      & f                       \\
MJD                    &                          \\
\hline\noalign{\smallskip}
53430--53800  & $96.1\pm3.2$   \\   
53800--54200  & $92.8\pm2.2$   \\   
54200--54600  & $90.0\pm2.5$   \\   
54600--54950  & $92.2\pm3.2$   \\   
54950--55250  & $95.5\pm2.5$   \\   
\hline\noalign{\smallskip}
%------------&-----------     \\
total time           &                           \\
 53430--55250  & $93.0\pm1.3$   \\   
\noalign{\smallskip}\hline
 \label{tab:tableA2}
\end{tabular}\end{center}
\end{table}
%Table A2 -----------------------------------------------------------------------------------

\end{document}